\documentclass[journal,12pt]{IEEEtran}

\usepackage{eurosym}
\usepackage{cite}
\usepackage{graphicx}
\usepackage{psfrag}
\usepackage{subfigure}
\usepackage{array}
\usepackage{mathtools}
\usepackage{amsfonts}
\usepackage{amssymb}
\usepackage{epsfig}
\usepackage{color}
\usepackage{t1enc}

\begin{document}

\title{Sinusoidally-Modulated Graphene Leaky-Wave Antenna for Electronic Beamscanning at THz}

\author{Marc Esquius-Morote, Juan Sebastian G\'{o}mez-D\'{i}az, and Julien Perruisseau-Carrier
\thanks{M. Esquius-Morote is with the Laboratoire d'Electromagn\'{e}tisme et d'Acoustique (LEMA), Ecole Polytechnique F\'{e}d\'{e}rale de Lausanne (EPFL), 1015 Lausanne, Switzerland (e-mail: marc.esquiusmorote@epfl.ch).}
\thanks{J. S. G\'{o}mez-D\'{i}az and J. Perruisseau-Carrier are with Adaptive MicroNano Wave Systems, LEMA/Nanolab, Ecole Polytechnique F\'{e}d\'{e}rale de Lausanne (EPFL), 1015 Lausanne,Switzerland (e-mails: juan-sebastian.gomez@epfl.ch, julien.perruisseau-carrier@epfl.ch)}
}

\maketitle

\begin{abstract}
This paper proposes the concept, analysis and design of a sinusoidally-modulated graphene leaky-wave antenna with beam scanning capabilities at a fixed frequency. The antenna operates at terahertz frequencies and is composed of a graphene sheet transferred onto a back-metallized substrate and a set of polysilicon DC gating pads located beneath it. In order to create a leaky-mode, the graphene surface reactance is sinusoidally-modulated via graphene's field effect by applying adequate DC bias voltages to the different gating pads. The pointing angle and leakage rate can be dynamically controlled by adjusting the applied voltages, providing versatile beamscanning capabilities. The proposed concept and achieved performance, computed using realistic material parameters, are extremely promising for beamscanning at THz frequencies, and could pave the way to graphene-based reconfigurable transceivers and sensors.

\end{abstract}


\IEEEpeerreviewmaketitle

\section{Introduction}
\label{sec:intro}

Graphene's unique electrical properties hold significant promises for the future implementation of integrated and reconfigurable (potentially all-graphene) terahertz transceivers and sensors. In this article, the term \textit{terahertz} is used to define the frequency range between 300\;GHz and 3\;THz \cite{Siegel02}. However, only few studies have considered the use of graphene in antenna applications. Initial works employed graphene as a component parasitic to radiation \cite{Dragoman10b, Mao12}, for instance as a switch element to chose among different states of a reconfigurable antenna. Then, the propagation of transverse-magnetic (TM) surface plasmon polaritons (SPPs) in graphene \cite{Jablan09, Sebas12_jap} was exploited in \cite{Tamagnone12_apl} to propose patch antennas in the THz band. This study was the first to consider graphene an actual antenna linking free space waves to a lumped source/detector, as needed in most communication and sensing scenarios. Several works have further studied the capabilities of graphene in antenna design \cite{Filter13, APS13} and even the integration of graphene in beam steering reflectarrays \cite{Carrasco13}.

On the other hand, different mechanisms have been proposed to excite SPPs in graphene structures, including the use of diffraction gratings \cite{Gao12,Peres12} or polaritonic crystals obtained by modulated graphene conductivity \cite{Bludov13}. However, the applications of THz transceivers which include chemical and biological remote sensing, image scanning, pico-cellular and intrasatellite communications, or high resolution imaging and tomography \cite{Federici05, rogalski2003infrared, lubecke98, Siegel02}, require antennas with specific capabilities in terms of beamscanning and directivity.

\begin{figure}[!t]
\centering
\includegraphics[width=0.95\columnwidth]{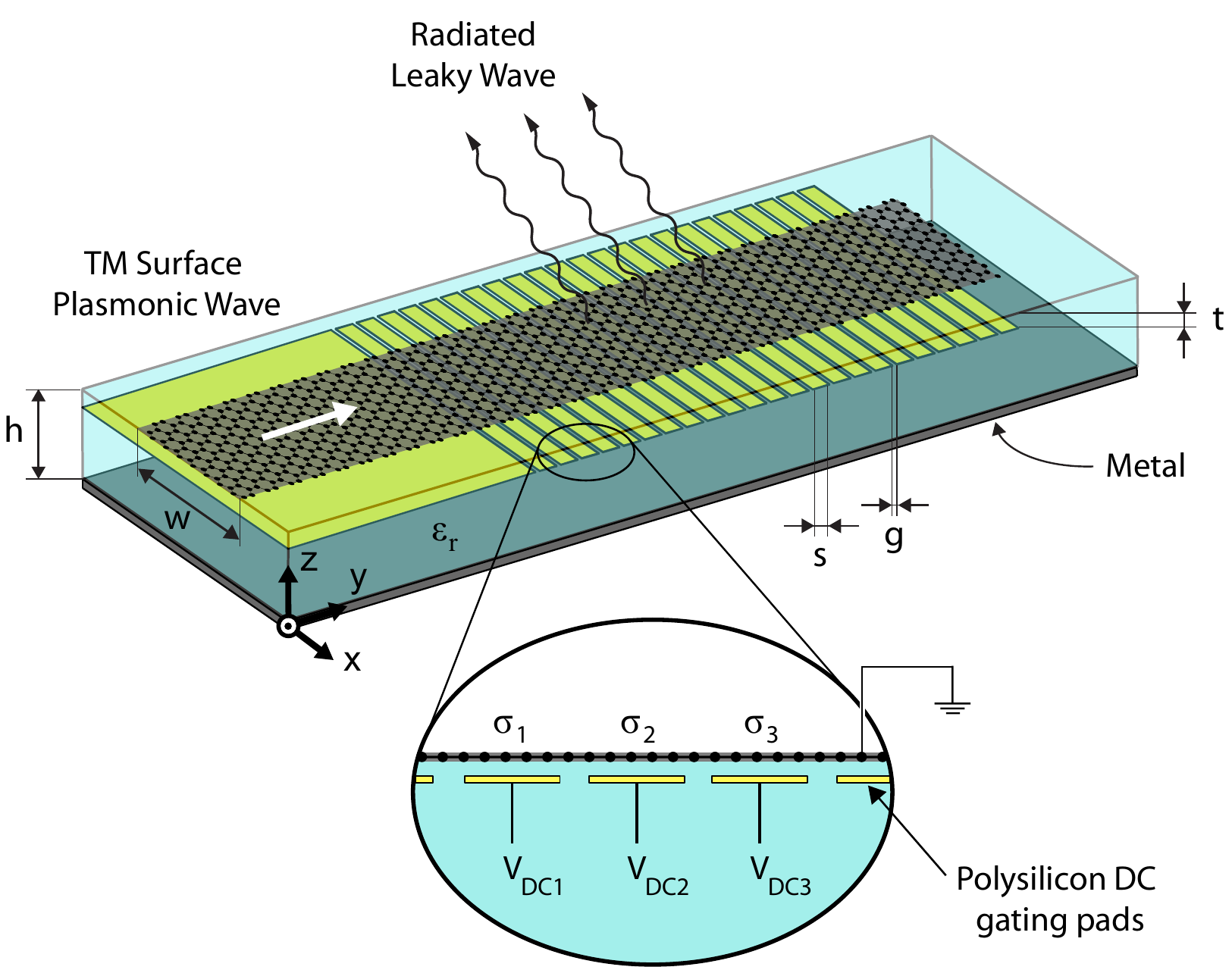}
\caption{Proposed sinusoidally-modulated reactance graphene surface operating as a leaky wave antenna. The polysilicon pads (yellow) are used to modify the graphene conductivity $\sigma_i$ as a function of the applied DC voltage $V_{DCi}$.}
\label{fig:GLWA}
\end{figure}

In this context, we propose the concept, analysis, and design of a graphene sinusoidally-modulated leaky-wave antenna (LWA) for electronic beamscanning in the THz band. The antenna is based on the principle of sinusoidally-modulated reactance surfaces to achieve leaky-wave radiation \cite{Oliner59, Grbic11_LWA_Sinusoidal, Maci11}. The proposed structure, shown in Fig.~\ref{fig:GLWA}, is composed of a graphene sheet transferred onto a back-metallized substrate and several polysilicon DC gating pads located beneath it. The graphene sheet supports the propagation of transverse-magnetic (TM) surface plasmon polaritons (SPPs) \cite{Sebas12_jap}, which can be controlled via the voltages applied to the different pads (note that the graphene sheet is connected to the source ground). Indeed, as illustrated in the inset of Fig.~\ref{fig:GLWA} and explained in more detail below, the application of a DC bias to the gating pads allows to control graphene's complex conductivity \cite{Geim2007}. This property is exploited to sinusoidally modulate the sheet surface reactance by applying adequate bias voltages to the different pads, thereby creating a leaky-wave mode \cite{Oliner59, Grbic11_LWA_Sinusoidal, Maci11}.

Moreover, antenna radiation features such as pointing angle ($\theta_0$) and leakage rate ($\alpha_{rad}$) can be dynamically controlled by modifying the applied voltages, thus allowing electronic beamscanning at a fixed operating frequency. This simple and fully integrated antenna structure is designed and analyzed using leaky-wave antenna theory \cite{Oliner_2007} and validated by full-wave simulations.

\section{Tunability of Graphene Conductivity}
\label{sec:SPP}
Graphene is a two-dimensional material composed of carbon atoms bonded in hexagonal structure. Its surface conductivity can be modeled using the well-known Kubo formalism \cite{Gusynin09}. In the low terahertz band and at room temperatures, interband contributions of graphene conductivity can safely be neglected \cite{Hanson09}. This allows one to describe graphene conductivity using only intraband contributions as
\begin{align}
\sigma=-j\frac{q_e^2K_BT}{\pi\hbar^2(\omega-j2\Gamma)}\left[\frac{\mu_c}{K_B T}+2\ln\left(e^{-\frac{\mu_c}{K_BT}}+1\right)\right],
\label{eq:Graphene_conductivity}
\end{align}
where $K_B$ is the Boltzmann's constant, $\hbar$ is the reduced Planck constant, $-q_e$ is the electron charge, $T$ is temperature, $\mu_c$ is graphene chemical potential, $\Gamma=1/(2\tau)$ is the electron scattering rate, and $\tau$ is the electron relaxation time.

Graphene conductivity can be tuned by applying a transverse electric field via a DC biased gating structure (see Fig.~\ref{fig:GLWA}). Assuming non-chemically doped graphene (i.e., $V_{Dirac} = 0$ \cite{Geim2007}), this field modifies the graphene carrier density $n_s$ as
\begin{equation}
C_{ox}V_{DC}=q_en_s,
\label{Capacitance_and_carrier_densitiy_relation}
\end{equation}
where $C_{ox}=\varepsilon_r\varepsilon_0/t$ is the gate capacitance and $V_{DC}$ is the applied DC bias field. Neglecting the quantum capacitance, i.e., assuming a relative thick gate oxide \cite{quantumC07}, the carrier density is related to graphene chemical potential $\mu_c$ as
\begin{equation}
n_s=\frac{2}{\pi\hbar^2v_f^2}\int_0^\infty\epsilon[f_d(\epsilon-\mu_c)-f_d(\epsilon+\mu_c)]\partial\epsilon,
\label{eq:Carrier_densitiy}
\end{equation}
where $\epsilon$ is energy, $v_f$ is the Fermi velocity ($\sim 10^8$ cm/s in graphene), and $f_d$ is the Fermi-Dirac distribution
\begin{equation}
f_d(\epsilon)=\left(e^{(\epsilon-|\mu_c|)/k_BT}+1\right)^{-1}.
\label{eq_Fermi_distribution}
\end{equation}
The chemical potential $\mu_c$ is accurately retrieved by numerically solving Eq.~(\ref{eq:Carrier_densitiy}). Hence, we can see from Eq. \eqref{eq:Graphene_conductivity} that the graphene conductivity $\sigma$ or surface impedance 1/$\sigma$ can be dynamically controlled by $V_{DC}$. This property can be used to create leaky-wave antennas with dynamic control as introduced in Section I and further detailed next.

\section{Sinusoidally-Modulated Reactance Surfaces}
\label{subsec:sinus_principle}
Electromagnetic propagation along sinusoidally-modulated reactance surfaces was theoretically investigated in the $50$s \cite{Oliner59}, but is still drawing significant interest as a powerful way to control radiation properties \cite{Maci11_Spiral_LWA,Grbic11_LWA_Sinusoidal}.
Surfaces with a dominant positive reactance impedance (such as graphene) allow the propagation of TM surface waves. If a modulation is then applied along the $y$ axis, the modal surface reactance $X_S$ can be expressed as:
\begin{equation}
 X_S = X_S'\left[1 + M\sin\left(\frac{2 \pi y}{p}\right)\right],
\label{eq:Xs}
\end{equation}
where $X_S'$ is the average surface reactance, $M$ the modulation index and $p$ the period of the sinusoid.

The interaction between surfaces waves and the reactance modulation produces a Bragg radiation effect thus creating leaky-wave radiation. The wavenumber $k_y$ of the fundamental space harmonic along the impedance modulated graphene sheet can be written as \cite{Oliner59}:
\begin{equation}
k_y =\beta_{spp}+\Delta\beta_{spp}-j(\alpha_{spp}+\alpha_{rad}),
\label{eq:k_y}
\end{equation}
where $\beta_{spp}$ is the propagation constant of the SPP on the unmodulated graphene sheet (i.e., with $M=0$), $\Delta\beta_{spp}$ a small variation in $\beta_{spp}$ due to the modulation, $\alpha_{spp}$ the attenuation constant due to losses and $\alpha_{rad}$ the attenuation constant due to energy leakage (leakage rate).

In this kind of periodic LWAs, the fundamental space harmonic is slow and usually the higher-order -1 space harmonic is used for radiation \cite{Oliner59}. In this case, the pointing angle $\theta_0$ of the radiated beam is approximated as \cite{Oliner_2007}
\begin{equation}
 \theta_0 = \arcsin\left(\frac{\beta_{-1}}{k_0}\right) \simeq \arcsin \left( \frac{\beta_{spp}}{k_0} - \frac{\lambda_0}{p} \right),
\label{eq:theta}
\end{equation}
where $\beta_{-1}$ is the propagation constant of the -1 space harmonic, $\lambda_0$ is a free space wavelength and $\Delta\beta_{spp}$ is assumed much smaller than $\beta_{spp}$ \cite{Maci11_Spiral_LWA}. Since the average surface reactance $X_S'$ mainly determines $\beta_{spp}$ and the modulation index $M$ mainly determines $\alpha_{rad}$, these LWAs allow for nearly independent control of the pointing angle and the beamwidth \cite{Grbic11_LWA_Sinusoidal}.

In the next Section, a graphene LWA based on this radiation principle is proposed taking into account realistic technological parameters for future implementation.

\section{Periodically Modulated Graphene LWA}
\label{sec:GLWA}
\subsection{Proposed Structure}
\label{subsec:proposed_LWA}

\begin{figure}[!t]
	\centering
	\includegraphics[width=0.98\columnwidth]{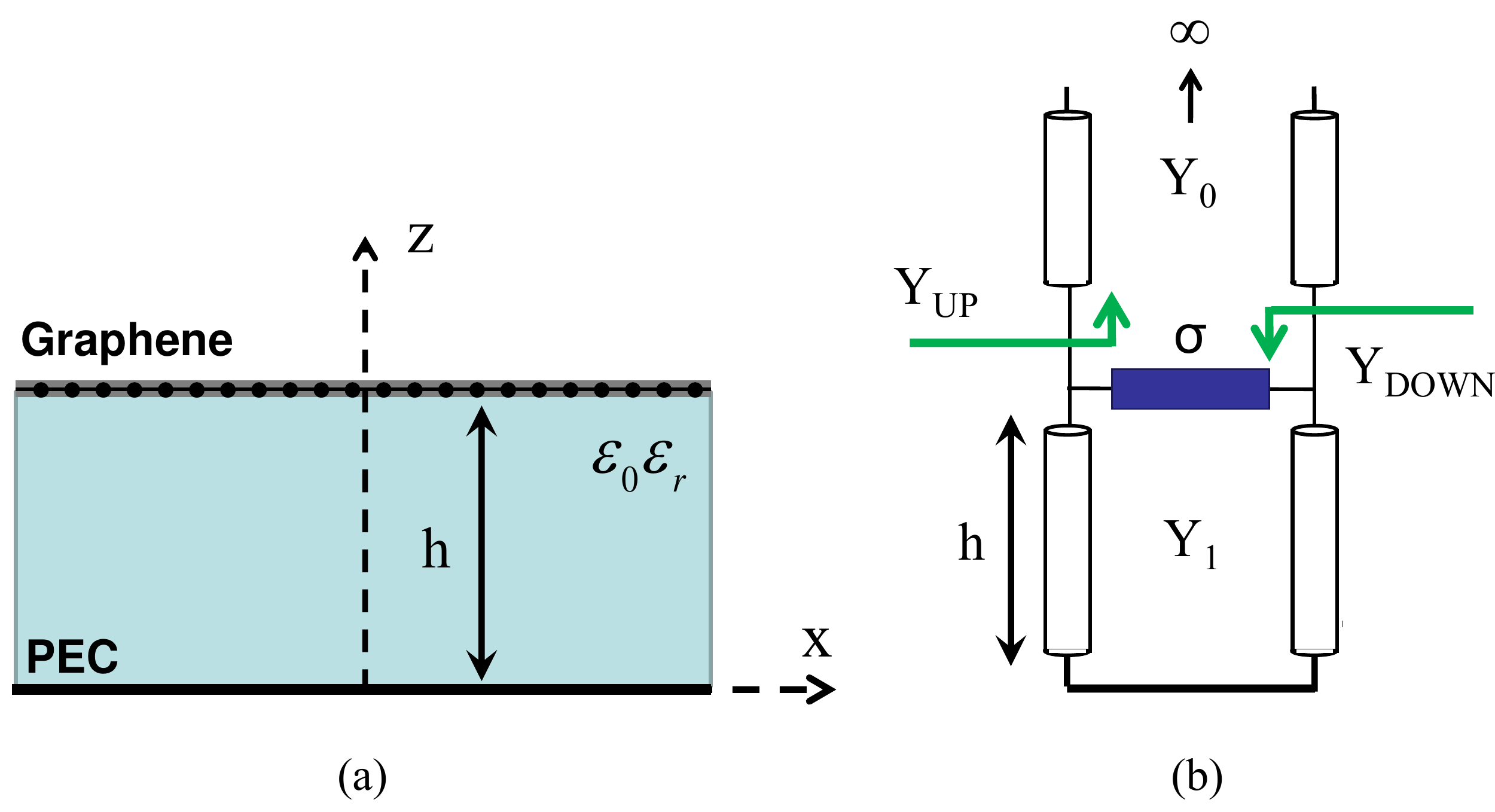}
	\caption{Schematic of graphene sheet transferred onto a back-metallized substrate (a) and its equivalent transverse network along the $z$-axis (b). The polysilicon pads have been neglected since they are extremely thin and with permittivity similar to the one of the Si02 substrate used.}
     \label{fig:TRE_figure}
\end{figure}

The proposed sinusoidally-modulated graphene LWA is shown in Fig.~\ref{fig:GLWA}. It consists of a graphene sheet transferred on a back-metallized substrate and several independent polysilicon DC gating pads beneath it. The surface impedance $Z_S = R_S + jX_S$ at the graphene position ($z=h$) determines the propagation characteristics of the SPPs, i.e., $k_{y,spp}=\beta_{spp}-j\alpha_{spp}$.

In an \textit{unmodulated} graphene sheet, the dispersion relation of the SPPs can be computed by applying a transverse resonance equation (TRE) to the equivalent circuit shown in Fig.~\ref{fig:TRE_figure} \cite{Itoh89}, where the width $w$ of the graphene sheet is considered infinite. Then, by enforcing $Y_{UP}+Y_{DOWN}=0$, where
\begin{align}
&Y_{UP}=\frac{\omega\varepsilon_{0}}{\pm\sqrt{k_0^2-k_{y,spp}^2}}, \\
&Y_{DOWN}=\sigma+\frac{\omega\varepsilon_{r}\varepsilon_{0}}{\pm\sqrt{\varepsilon_{r}k_0^2-k_{y,spp}^2}}\coth(\pm \sqrt{\varepsilon_{r}k_0^2-k_{y,spp}^2} h),
\label{eq:Equivalent_admittances}
\end{align}
%
the desired dispersion relation is obtained as
\begin{align}
\frac{\omega\varepsilon_{0}}{\pm\sqrt{k_0^2-k_{y,spp}^2}}&+\nonumber \\\frac{\omega\varepsilon_{r}\varepsilon_{0}}{\pm\sqrt{\varepsilon_{r}k_0^2-k_{y,spp}^2}}&\coth\left(\pm \sqrt{\varepsilon_{r}k_0^2-k_{y,spp}^2} h\right)=-\sigma.
\label{eq:Dispersion_TM_general}
\end{align}
In the above expressions, $\varepsilon_{0}$ is the vacuum permittivity, $\varepsilon_{r}$ is the dielectric permittivity, $k_0=\omega/c$ is the free space wavenumber and $h$ is the dielectric thickness. The surface impedance $Z_S$ is then computed as $1/Y_{DOWN}$. It can be shown that the polysilicon pads can be safely neglected since they are extremely thin ($\sim 100\;$nm) and with a relative permittivity $\varepsilon_r \simeq 3$ \cite{Pryputniewicz2002} similar to the one of the SiO$_2$ substrate used.

\begin{figure}[!t]
\centering
\includegraphics[width=\columnwidth]{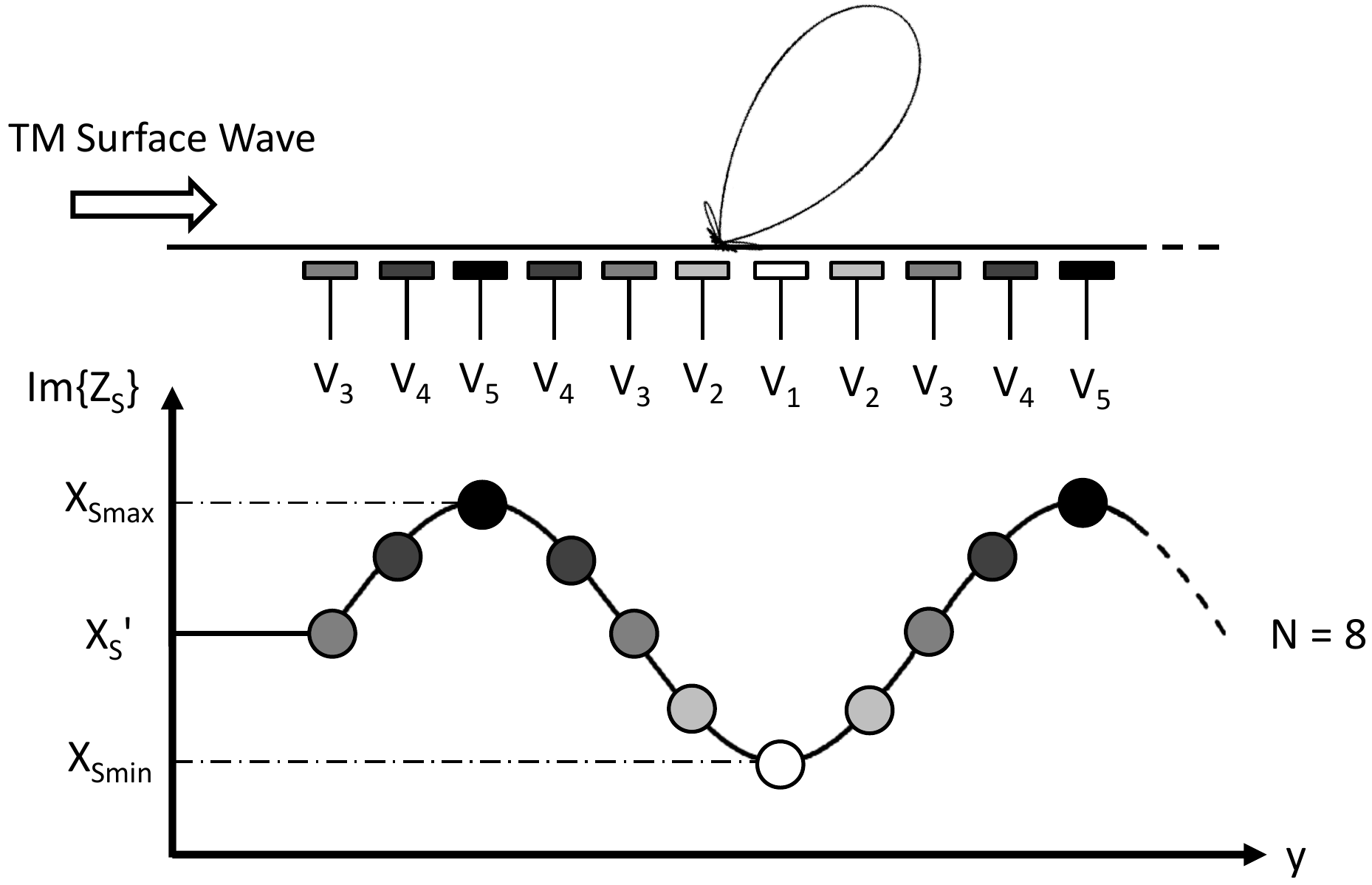}
\caption{Schematic representation of the relationship between the DC bias voltage $V_{DC}$ and graphene reactance $X_S$.}
\label{fig:reconf}
\end{figure}

As explained in Section \ref{sec:SPP}, in order to modulate the surface impedance $Z_S$, the graphene conductivity $\sigma$ is modified by applying adequate bias voltages $V_{DC}$ to the different pads. This allows one to generate the desired periodic reactance modulation. Here, one modulation period $p$ is sampled in $N$ points according to the number of gating pads used, namely, $p=N(s+g)$, where $s$ is the pad length and $g$ the distance between pads (see Fig.~\ref{fig:reconf}). The dimensions $s$ and $g$ should be chosen such that $g\ll s$ in order to guarantee a fairly constant perpendicular electrostastic field at the graphene section above each polysilicon pad. Therefore, $p$ can be dynamically controlled according to the periodicity imposed by the different $V_{DC}$. Consequently, from Eq. \eqref{eq:theta}, the proposed structure offers electronic beamscanning at a fixed frequency. This is a novel property absent when usual techniques such as the use of sub-wavelengths printed elements or dielectric slabs with variable thickness \cite{Sievenpiper10_Holography,Grbic11_LWA_Sinusoidal,Maci11_Spiral_LWA} are employed to implement sinusoidally-modulated LWA.

The impedance-modulated graphene sheet of Fig.~\ref{fig:GLWA} may radiate towards both the upper and lower half-spaces depending on the substrate permittivity $\varepsilon_r$. The metallic plane is then used as a reflector and one should choose the substrate thickness $h$ such that the main and the reflected beams add in phase. In addition, it is important to remark that, although SPPs on graphene sheets are confined into the layer \cite{Sebas12_jap}, the presence of a metallic plane at a distance $h$ from the graphene sheet might affect their propagation characteristics.  This is particularly important when dealing with this type of LWA due to their high dispersive nature, where a small variation on $\beta_{spp}$ can considerably tilt the pointing angle $\theta_0$. Here, this phenomenon is taken into account in the equivalent circuit of Fig. \ref{fig:TRE_figure} and is illustrated in the Section \ref{sec:Design_example}.

\subsection{Design Strategy}
The design flow of the proposed graphene LWA must take into account the technological limitations of current graphene fabrication processes. Therefore, given the parameters of a graphene sample ($\tau$, $T$, $t$, $V_{DC}$), the substrate ($\varepsilon_r$), the frequency of operation $f_0$ and the pointing angle $\theta_0$, the design steps are the following:

\begin{enumerate}
\item Given $\theta_0$, the needed propagation constant of the -1 space harmonic $\beta_{-1}$ is computed with Eq. \eqref{eq:theta}.

\item Knowing the substrate and the graphene properties, the propagation constant $\beta_{spp}$ of the SPPs is obtained. First, the available values of $\mu_c$ are found with Eqs. \eqref{Capacitance_and_carrier_densitiy_relation} and \eqref{eq:Carrier_densitiy} as a function of $t$ and the range of $V_{DC}$. Then, graphene conductivity $\sigma$ is computed with Eq. \eqref{eq:Graphene_conductivity} and, in turn, the wavenumber $k_{y,spp}$ of the SPPs with Eq. \eqref{eq:Dispersion_TM_general}.

\item The required modulation period $p$ is then computed as $p = 2\pi/(\beta_{spp}-\beta_{-1})$.

\item The next step is to find the possible values of the modulation index $M$ that can be obtained with the graphene sample. First, the surface impedances $Z_S = R_S + jX_S$ are computed as a function of $\mu_c$. Then, the desired range of reactances ($X_{S_{min}} - X_{S_{max}}$ in Fig.~\ref{fig:reconf}) is chosen, thus defining the value of $M$.

\item The leakage factor $\alpha_{rad}$ can now be computed as described in \cite{Oliner59}.

\item If beamscanning around $\theta_0$ is desired, the new periodicities around $p$ are computed with \eqref{eq:theta}. The dimensions of the gating pads $s$ and $g$ and the different values of $N$ are chosen in order to synthesize the required periods.

\item Knowing the scanning range, the substrate thickness $h$ is then chosen so that there is a constructive interference between the main and reflected beams in all the different angles.

\item Finally, the complex propagation constant $k_y$ is determined by Eq. \eqref{eq:k_y} and the radiation pattern of the LWA can be easily computed using standard techniques \cite{Oliner_2007}.

\end{enumerate}

\section{Design Example}
\label{sec:Design_example}

In this example we consider a SiO$_2$ substrate of permittivity $\varepsilon_r = 3.8$ \cite{Naftaly07_THz_TDS}. We assume graphene with a relaxation time $\tau=1$~ps, temperature $T=300^{\circ}$~K and design the antenna for operation at $f_0=2$~THz. The value of $\tau$ was estimated from the measured impurity-limited DC graphene mobility on boron nitride \cite{Dean10} of $\mu \simeq 60000 cm^2/(V~s)$, which leads to $\tau = \mu E_F /(q_ev_f^2) \simeq 1.2$~ps for a Fermi level of $E_F = 0.2$~eV. Note that even higher mobilities have been observed in high quality suspended graphene \cite{Bolotin08}.

The width $w$ of the graphene strip is chosen to be electrically very large (200\;$\mu$m) in order to support the propagation of SPPs with the same characteristics as in 2D infinite sheets \cite{Nikitin11} thus allowing to use the simplified model of Fig. \ref{fig:TRE_figure}.

\begin{figure}[!t]
\centering
\includegraphics[width=0.95\columnwidth]{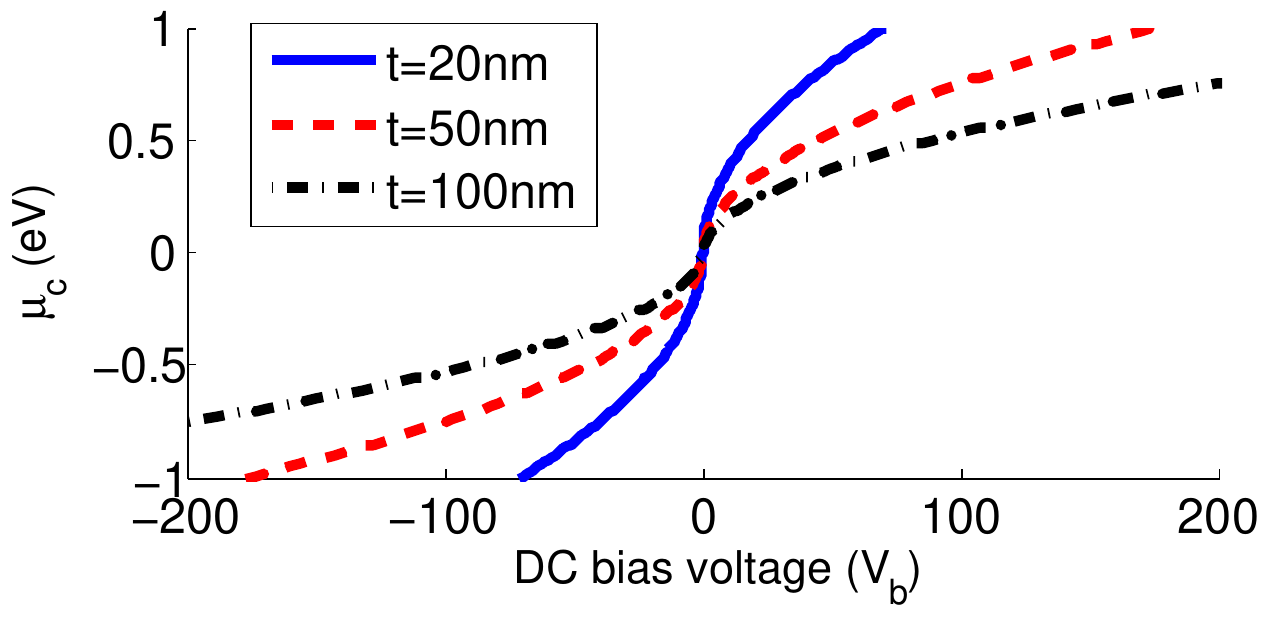}
\caption{Graphene chemical potential $\mu_c$ as a function of the DC bias voltage $V_{DC}$. Parameters: $\tau=1$~ps, $T=300^{\circ}$~K, $f_0$=2~THz and $\varepsilon_r=3.8$.}
\label{fig:voltages}
\end{figure}

The graphene chemical potential $\mu_c$ as a function of the DC bias voltage $V_{DC}$, computed using Eqs. \eqref{Capacitance_and_carrier_densitiy_relation} and \eqref{eq:Carrier_densitiy}, is shown in Fig. \ref{fig:voltages} for different values of $t$. A distance $t = 20$\;nm is chosen which offers values of $\mu_c$ until 0.8\;eV using $V_{DC}$ values below 45\;V.

\begin{figure}[!t]
\centering
\includegraphics[width=0.95\columnwidth]{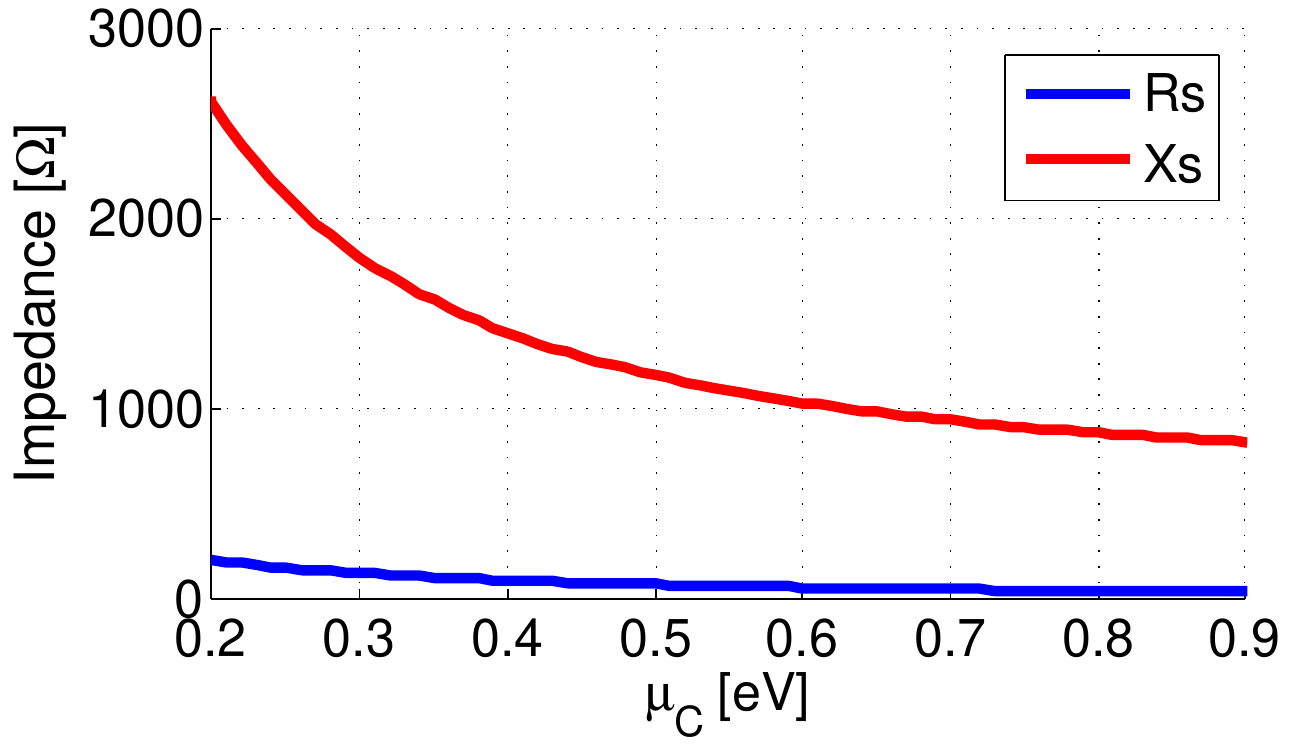}
\caption{Surface resistance $R_S$ and reactance $X_S$ at the graphene sheet (position $z=h$ in Fig.~\ref{fig:TRE_figure}) as a function of the chemical potential $\mu_c$. Parameters: $\tau=1$~ps, $T=300^{\circ}$~K, $f_0$=2~THz and $\varepsilon_r=3.8$.}
\label{fig:impedance}
\end{figure}

The surface impedance $Z_S$ is plotted in Fig.~\ref{fig:impedance} as a function of $\mu_c$. It is observed that, while the graphene resistance $R_S$ remains fairly constant, the reactance $X_S$ can be considerably modified.
Under these conditions, an average surface reactance $X_S'=1302~\Omega/\Box$ and a modulation index $M = 0.35$ are chosen. According to Fig. \ref{fig:voltages} and \ref{fig:impedance}, this requires a range of chemical potentials between $0.3$ and $0.8$~eV, i.e., a range of DC voltages between 6.4\;V and 45\;V.

\begin{figure}[!t]
  \begin{center}
    \subfigure[]{\includegraphics[width=0.9\columnwidth]{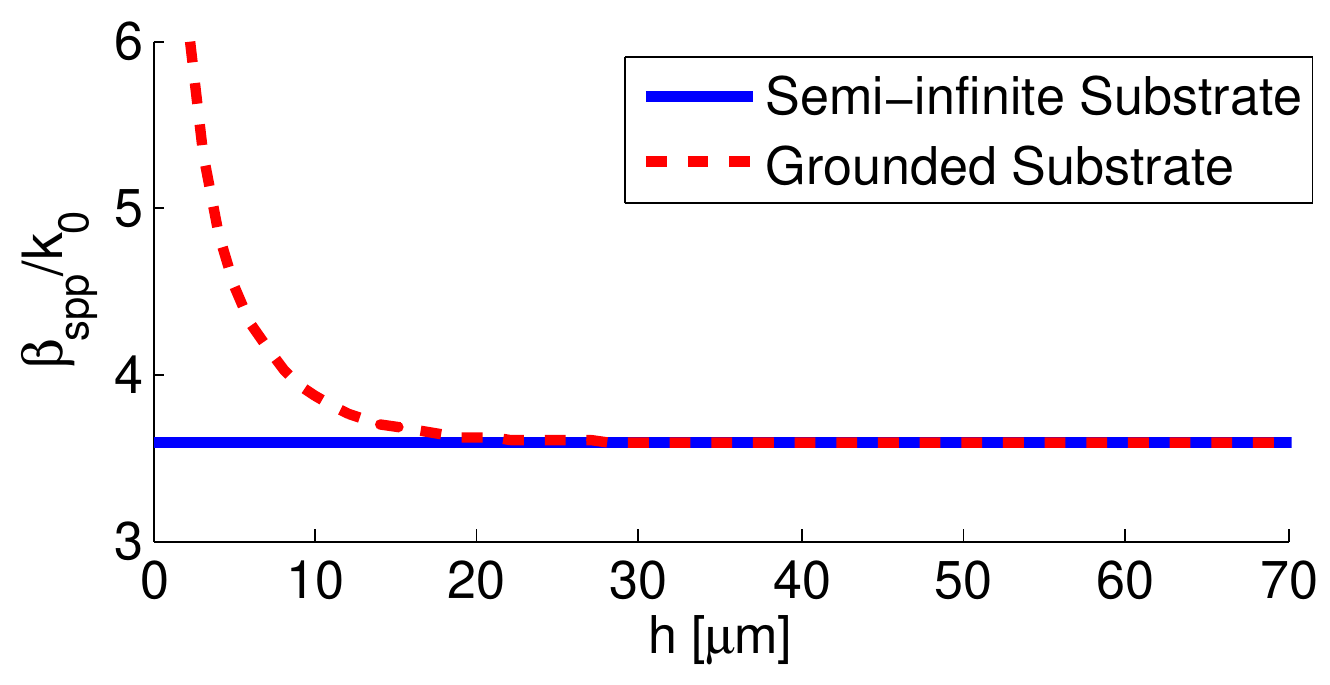}\label{fig:BSW_h}} \\
  	\subfigure[]{\includegraphics[width=0.98\columnwidth]{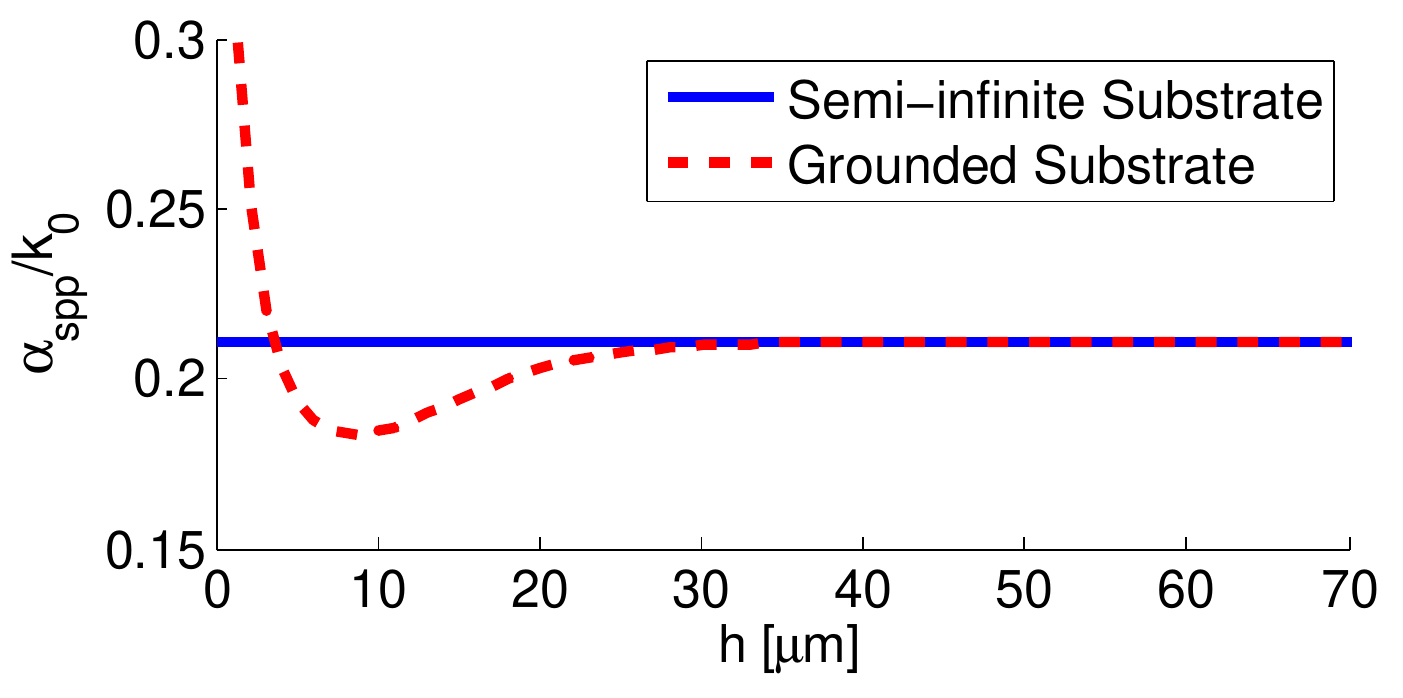}\label{fig:alphaL_h}}
  \end{center}
	\caption{Normalized phase constant $\beta_{spp}/k_0$ (a) and dissipation losses $\alpha_{spp}/k_0$ (b) of a SPP on the structure shown in Fig.~\ref{fig:TRE_figure} as a function of the substrate thickness $h$. Parameters: $\tau=1$~ps, $T=300^{\circ}$~K, $f_0$=2~THz, $V_{DC}=13.5$~V and $\varepsilon_r=3.8$.}
     \label{fig:h_effect}
\end{figure}

The voltage $V_{DC}$ needed for the desired $X_S'$ is 13.5\;V, i.e., $\mu_c = 0.436$\;eV. This corresponds to $\beta_{spp}/k_0 = 3.59$ and $\alpha_{spp}/k_0 = 0.21$. In order to illustrate the effect of the metallic plane on the propagation of SPPs, $\beta_{spp}$ and $\alpha_{spp}$ are plotted in Fig.~\ref{fig:h_effect} as a function of $h$. The solid blue lines indicate the values obtained in the case of a graphene sheet on a semi-infinite substrate. It can be seen that for $h > 30\mu$m the effect of the metallic plane is negligible and thus the plasmon is indeed highly confined to the graphene layer.

\begin{figure}[!t]
\centering
\includegraphics[width=0.9\columnwidth]{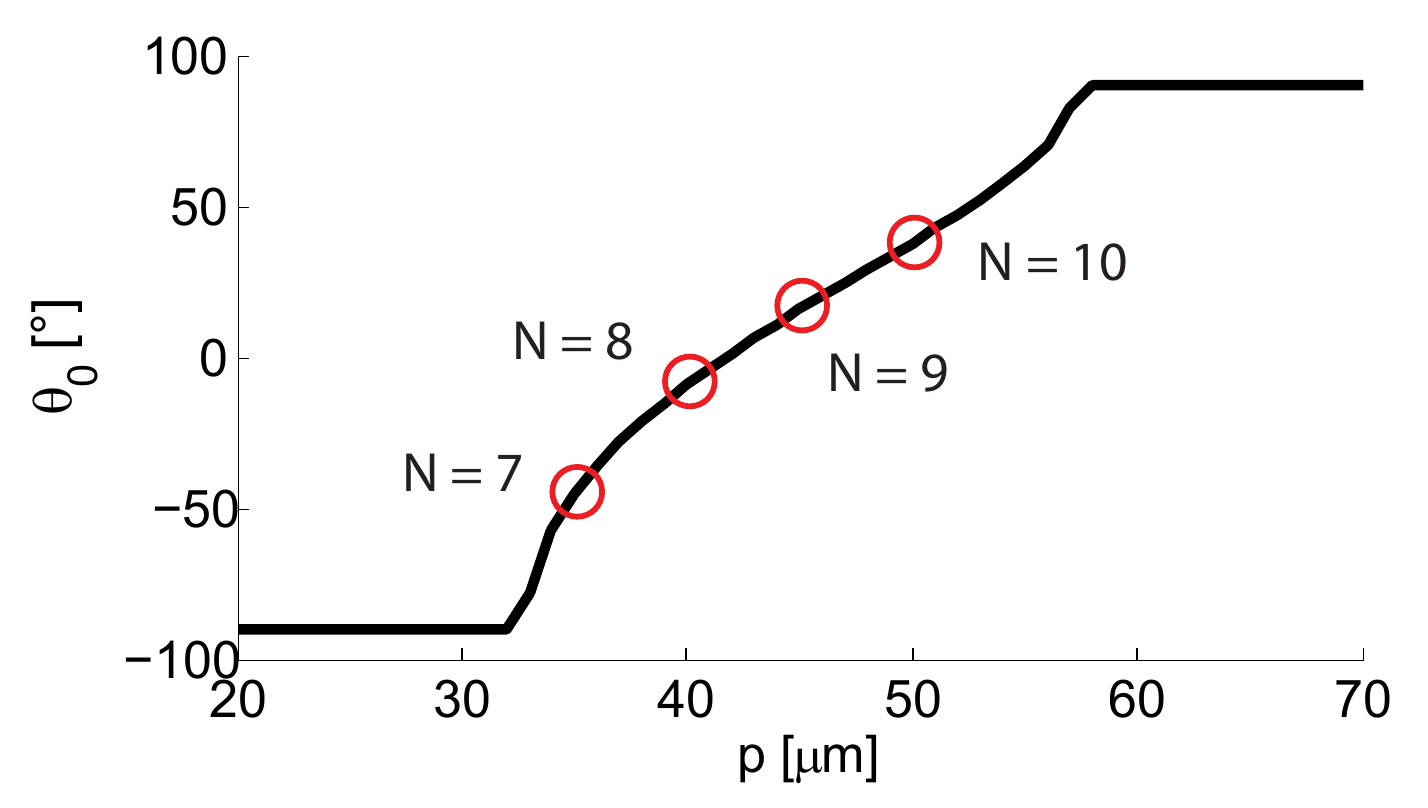}
\caption{Poiting angle $\theta_0$ as a function of the period $p$. N is the number of polysilicon pads used to define one modulation period $p$. Parameters: $\tau=1$~ps, $V_{DC}=13.5$~V, $T=300^{\circ}$~K, $f_0$=2~THz, $\varepsilon_r=3.8$, $s = 4.8\mu$m and $g = 0.2\mu$m.}
\label{fig:theta_period}
\end{figure}

\begin{figure}[!t]
\centering
\includegraphics[width=\columnwidth]{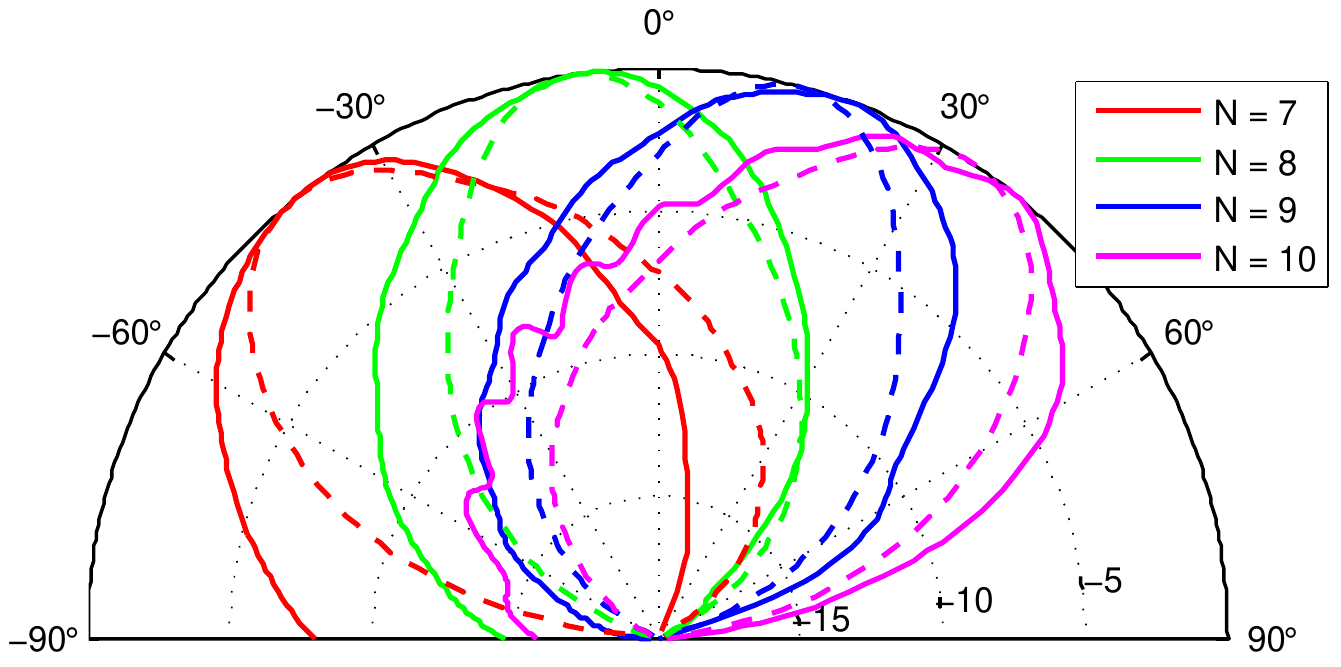}
\caption{Radiation patterns (ZY plane) for different values of N. Solid lines are obtained with HFSS v14 and dashed lines with LWA theory \cite{Oliner_2007}. Parameters: $\tau=1$~ps, $T=300^{\circ}$~K, $f_0$=2~THz, $V_{DC}=6.5 - 45$~V, $\varepsilon_r=3.8$, $s = 4.8\mu$m and $g = 0.2\mu$m.}
\label{fig:patterns2}
\end{figure}

Once $\beta_{spp}$ is known, Eq. \eqref{eq:theta} allows determining that the -1 space harmonic is in the fast-wave radiation region for values of $p$ comprised between 33 and 57\;$\mu$m (see Fig. \ref{fig:theta_period}). Depending on the size of the polysilicon pads, a different number of scanning beams can be obtained. Although smaller pads offer the possibility to generate more beams, the complexity of the antenna also increases. As a trade-off, the dimensions of the pads are here chosen as $s = 4.8\;\mu$m and $g = 0.2\;\mu$m, which satisfy $g\ll s$ and can be manufactured with current technology. This configuration can generate 4 different beams at $-45.4$, $-9.3$, $15.4$ and $37.5^{\circ}$ by varying $N$ from 7 to 10 (see Fig. \ref{fig:theta_period}). The radiation patterns obtained with LWA theory and full-wave simulations for these values of $N$ are plotted in Fig. \ref{fig:patterns2}.

In this case, a substrate thickness $h=60\mu$m is chosen. A parametric study was performed to determine the optimum value of $h$ so that there is a constructive interference between the main and reflected beams in all the different scanning angles. For the reader's convenience, a summary of the different parameters used in this design is presented in Tab. \ref{tab:design_param}.

\begin{table}[!t]
\centering
\renewcommand{\arraystretch}{1.4}
\caption{Parameters used in the proposed design example.}
\label{tab:design_param}
\begin{tabular}{|c|c|c|c|c|c|c|c|}
  \hline
  \multicolumn{3}{|c|}{\textbf{Graphene}} & \multicolumn{2}{|c|}{\textbf{Substrate}} & \multicolumn{3}{|c|}{\textbf{Polysilicon pads}} \\
  \hline
  $\tau$ & T & w & $\varepsilon_r$ & h & t & s & g \\ \hline
  1ps & 300$^{\circ}$K & 200$\mu$m & 3.8 & 60$\mu$m & 20nm & 4.8$\mu$m & 0.2$\mu$m \\
  \hline
\end{tabular}
\end{table}

In the scanning region, the proposed LWA presents a leakage factor $\alpha_{rad}/k_0\simeq 0.025$ and dissipation losses  $\alpha_{spp}/k_0\simeq 0.21$. The effective length of the antenna for a 95$\%$ of dissipated power is $L_e = 3/(2(\alpha_{rad}+\alpha_{spp}))$ \cite{Oliner_2007}. The radiation efficiency $\eta_{rad}$ of LWAs with non-negligible dissipation losses is given by \cite{GomezTornero12}:

\begin{equation}
\eta_{rad} = \frac{\alpha_{rad}}{\alpha_{rad}+\alpha_{spp}}\left(1-e^{-2(\alpha_{rad}+\alpha_{spp})L_e}\right).
\label{eq:radeff}
\end{equation}

In our case, $L_e \simeq \lambda_0$ and $\eta_{rad} \simeq 11 \%$. The radiation efficiency could be further increased by using a wider range of $V_{DC}$ to achieve higher values of $M$, e.g., $M=0.5$ leads to $\eta_{rad} \simeq 19 \%$.

\begin{figure}[!t]
\centering
\includegraphics[width=0.85\columnwidth]{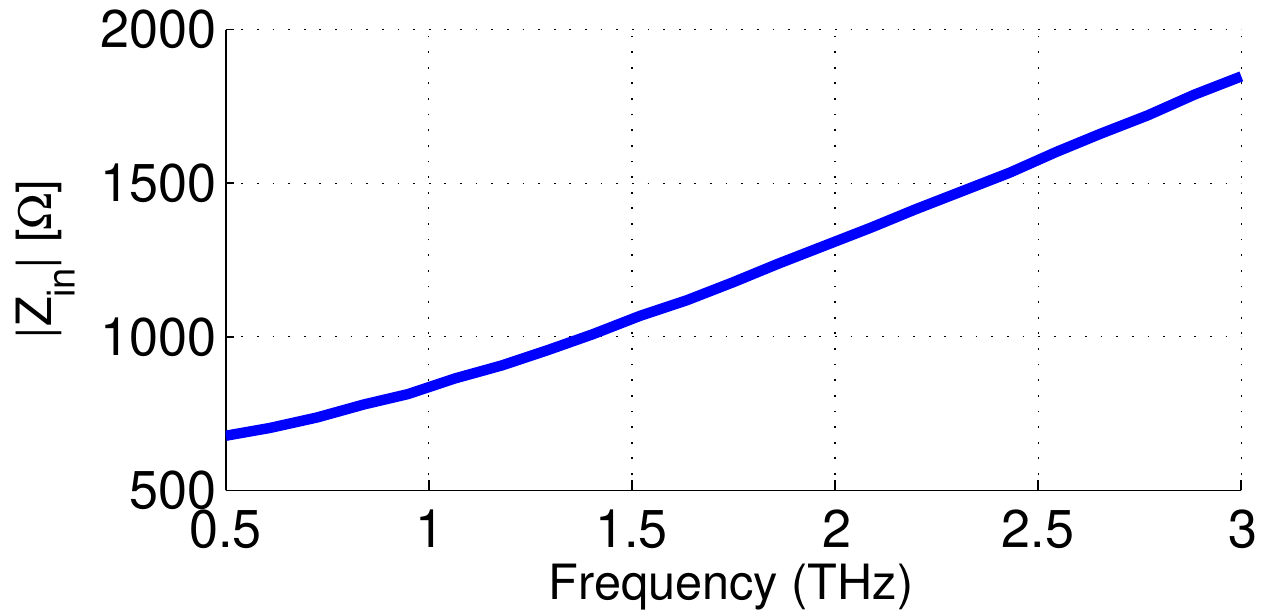}
\caption{Antenna input impedance vs frequency. Parameters: $\tau=1$~ps, $T=300^{\circ}$~K, $\varepsilon_r=3.8$, $s = 4.8\mu$m and $g = 0.2\mu$m.}
\label{fig:inputZ}
\end{figure}

The antenna input impedance as a function of the frequency is shown in Fig. \ref{fig:inputZ}. Note that presenting high impedance values is a very desirable feature for minimizing return loss when connecting the antenna to a photomixer, since these devices present a high input impedance with values in the range of K$\Omega$ \cite{Woo10,Yi11}.

\section{Practical Considerations}
\label{sec:practical}

This section briefly describe practical consideration related to the antenna's performance and fabrication.

\begin{figure}[!t]
\centering
\includegraphics[width=\columnwidth]{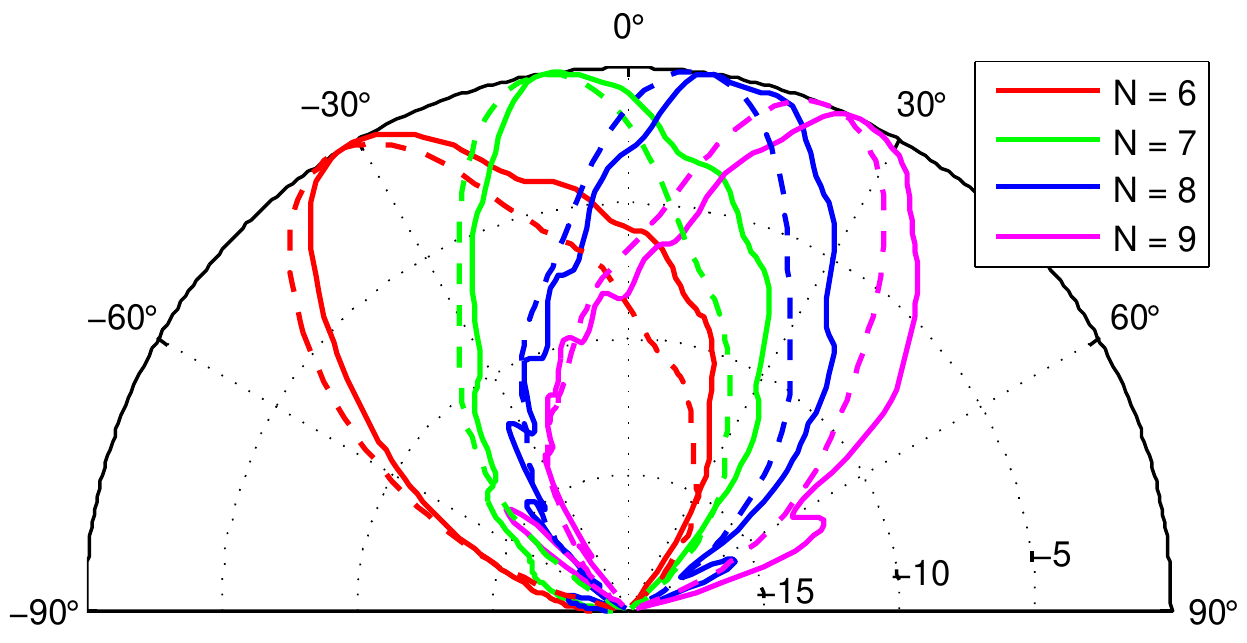}
\caption{Radiation patterns (ZY plane) for different values of N. Solid lines are obtained with HFSS v14 and dashed lines with LWA theory \cite{Oliner_2007}. Parameters: $\tau=1$~ps, $T=300^{\circ}$~K, $f_0$=2~THz, $V_{DC}=6.5 - 45$~V, $\varepsilon_r=1.8$, $s = 8.8\mu$m and $g = 0.2\mu$m.}
\label{fig:patterns}
\end{figure}

First, the antenna directivity is increased when using lower permittivity substrates. Indeed, reducing $\varepsilon_r$ reduces $\beta_{spp}$ and $\alpha_{spp}$ and thus the effective length of the antenna is increased. For instance, using the same graphene sheet as in Section \ref{sec:Design_example} on a substrate with $\varepsilon_r = 1.8$ implies $\beta_{spp}/k_0 = 2.23$, $\alpha_{spp}/k_0\simeq 0.117$ and $\alpha_{rad}/k_0\simeq 0.02$. Obviously, since $\beta_{spp}$ is reduced, the values of $p$ needed to scan with the -1 harmonic are larger. In this case, using pads of $s = 8.8\mu$m and $g = 0.2\mu$m and values of $N$ between 6 and 9 generates the beams shown in Fig. \ref{fig:patterns}. The effective length of the antenna is now $1.74\lambda_0$ allowing for higher directive beams and a radiation efficiency around 15$\%$. The drawback of this configuration is that low permittivity substrates are difficult to implement and a higher number of polysilicon pads is required.

Second, the radiation efficiency of the antenna always depends on the graphene quality, determined by the relaxation time $\tau$. As $\tau$ increases, $\alpha_{spp}$ decreases and so do the losses. Measured values of $\tau=1.0$\;ps have been used in this work to obtain the radiation efficiencies. Note that, even though these values are not very high, they are within the state-of-the-art of THz antennas \cite{Siegel02}. Moreover, most THz antennas do not include reconfiguration capabilities.

Third, the possible operation frequencies of the proposed structure are limited by graphene quality and the intrinsic behaviour of TM SPPs. On the one hand, the minimum operating frequency is within the low terahertz band ($\simeq$ 0.5\;THz) where graphene losses start to considerably increase and the surface reactance cannot be widely tuned by applying reduced DC voltages. On the other hand, the maximum operating frequency is predicted to be at few tenths of THz. At these frequencies, the plasmons are extremely confined at the graphene sheet and the dimensions of the polysilicon pads become really small to be implemented with current manufacturing processes. In addition, at very high frequencies (usually starting from 8 or 10\;THz), the interband contributions of graphene conductivity are no longer negligible \cite{Hanson09} and further theoretical studies would be necessary.

Concerning the fabrication process of the proposed device, the following procedure could be carried out. A back-metallized bare Si p-type wafer can be employed as sacrificial substrate, covered by two dielectric layers (conducting polysilicon and SiO2) by using a deposition process. Before depositing the SiO2, the polysilicon must be patterned using e-beam lithography. Then, graphene is transferred onto the stacks. Metalling contact must be defined at the edge of the polysilicon pads by using optical lithography, followed by the evaporation of gold with a chromium adhesive layer and a lift-off process \cite{pads12}.

It is also important to mention that the fringing electrostatic fields between consecutive polysilicon pads can be neglected and, therefore, they do not perturb the conductivity profile along the z-axis on graphene. First, it is shown in \cite{fringing_13} that a finite-width gating pad imposes a \textit{soft} boundary condition on the graphene sheet, leading to a smooth conductivity profile. Second, in order to provide a sinusoidal modulation of graphene's surface reactance, consecutive pads are DC biased with relatively similar voltages (as graphically illustrated in Fig. \ref{fig:reconf}) thus reducing even more the possible influence of the fringing fields.

Finally, it is worth mentioning that no power handling issues are foreseen in the proposed antenna structure. Similar mono-layer configurations have been studied and measured exhibiting really good performances in terms of thermal conductivity  \cite{08_Power_Graphene}.

\section{Conclusions}

A graphene leaky-wave antenna which allows electronic beamscanning at a single frequency has been proposed. Its radiation principle is based on sinusoidally-modulated reactance surfaces which can be easily implemented using a graphene sheet thanks to its tunable characteristics when applying electric field biasing.

This novel antenna concept offers unprecedented performances at the THz band such as electronic control of several  radiation characteristics preserving a radiation efficiency above 10$\%$. The main limitations in its design are imposed by graphene properties (such as the relaxation time $\tau$), the range of feasible voltages that can be applied to the gating pads, and the availability of substrates in the THz band.

The overall performances of this simple antenna structure are very promising for its integration in future all-graphene reconfigurable THz transceivers and sensors.


\section*{Acknowledgement}
This work was supported by the Swiss National Science Foundation (SNSF) under grant $133583$ and by the EU FP$7$ Marie-Curie IEF grant "Marconi", with ref. $300966$. The authors wish to thank Prof.~J.~R.~Mosig (\'Ecole Polytechnique F\'ed\'erale de Lausanne, Switzerland), Dr. Mar\'{i}a Garc\'{i}a Vigueras (\'Ecole Polytechnique F\'ed\'erale de Lausanne, Switzerland) and Prof.~A.~Grbic (University of Michigan) for fruitful discussion.



\begin{thebibliography}{10}
\providecommand{\url}[1]{#1}
\csname url@samestyle\endcsname
\providecommand{\newblock}{\relax}
\providecommand{\bibinfo}[2]{#2}
\providecommand{\BIBentrySTDinterwordspacing}{\spaceskip=0pt\relax}
\providecommand{\BIBentryALTinterwordstretchfactor}{4}
\providecommand{\BIBentryALTinterwordspacing}{\spaceskip=\fontdimen2\font plus
\BIBentryALTinterwordstretchfactor\fontdimen3\font minus
  \fontdimen4\font\relax}
\providecommand{\BIBforeignlanguage}[2]{{%
\expandafter\ifx\csname l@#1\endcsname\relax
\typeout{** WARNING: IEEEtran.bst: No hyphenation pattern has been}%
\typeout{** loaded for the language `#1'. Using the pattern for}%
\typeout{** the default language instead.}%
\else
\language=\csname l@#1\endcsname
\fi
#2}}
\providecommand{\BIBdecl}{\relax}
\BIBdecl

\bibitem{Siegel02}
P.~H. Siegel, ``Terahertz technology,'' \emph{IEEE Transactions on Microwave
  Theory and Techniques}, vol.~50, pp. 910--928, 2002.

\bibitem{Dragoman10b}
M.~Dragoman, A.~A. Muller, D.~Dragoman, F.~Coccetti, and R.~Plana, ``Terahertz
  antenna based on graphene,'' \emph{Journal of Applied Physics}, vol. 107, p.
  104313, 2010.

\bibitem{Mao12}
Y.~Huang, L.~S. Wu, M.~Tang, and J.~Mao, ``Design of a beam reconfigurable thz
  antenna with graphene-based switchable high-impedance surface,'' \emph{IEEE
  Trans. on Nanotechnol.}, vol.~11, no.~4, pp. 836--842, 2012.

\bibitem{Jablan09}
M.~Jablan, H.~Buljan, and M.~Soljacic, ``Plasmonics in graphene at infrared
  frequencies,'' \emph{Physical review B}, vol.~80, p. 245435, 2009.

\bibitem{Sebas12_jap}
J.~S. G\'omez-D\'iaz and J.~Perruisseau-Carrier, ``Propagation of hybrid
  transverse magnetic-transverse electric plasmons on magnetically-biased
  graphene sheets,'' \emph{Journal of Applied Physics}, vol. 112, p. 124906,
  2012.

\bibitem{Tamagnone12_apl}
M.~Tamagnone, J.~S. G\'omez-D\'iaz, J.~R. Mosig, and J.~Perruisseau-Carrier,
  ``Reconfigurable thz plasmonic antenna concept using a graphene stack,''
  \emph{Applied Physics Letters}, vol. 101, p. 214102, 2012.

\bibitem{Filter13}
R.~Filter, M.~Farhat, M.~Steglich, R.~Alaee, C.~Rockstuhl, and F.~Lederer,
  ``Tunable graphene antennas for selective enhancement of thz-emission,''
  \emph{Optics Express}, vol.~21, pp. 3737--3745, 2013.

\bibitem{APS13}
J.~Perruisseau-Carrier, M.~Tamagnone, J.~S. G\'omez-D\'iaz, M.~Esquius-Morote,
  and J.~R. Mosig, ``Resonant and leaky-wave reconfigurable antennas based on
  graphene plasmonics,'' \emph{IEEE Antennas Propag. Society International
  Symposium (APSURSI)}, 2013.

\bibitem{Carrasco13}
E.~Carrasco and J.~Perruisseau-Carrier, ``Tunable graphene reflective cells for
  thz reflectarrays and generalized law of reflection,'' \emph{Appl. Phys.
  Lett.}, vol. 102, pp. 104--103, 2013.

\bibitem{Gao12}
W.~Gao, J.~Shu, C.~Qiu, and Q.~Xu, ``Excitation of plasmonic waves in graphene
  by guided-mode resonances,'' \emph{ACS nano}, vol.~6, pp. 7806--7813, 2012.

\bibitem{Peres12}
N.~M. Peres, Y.~V. Bludox, A.~Ferreira, and M.~I. Vasilevskiy, ``Exact solution
  for square-wave grating covered with graphene: Surface plasmon-polaritons in
  the thz range,'' \emph{Arxiv preprint}, vol. 1211.6358v1, 2012.

\bibitem{Bludov13}
Y.~V. Bludox, A.~Ferreira, N.~M. Peres, and M.~I. Vasilevskiy, ``A primer on
  surface plasmon-polaritons in graphene,'' \emph{Arxiv preprint}, vol.
  1302.2317v1, 2013.

\bibitem{Federici05}
B.~S. J.~F.~Federici, F.~Huang, D.~Gary, R.~Barat, F.~Oliveira, and D.~Zimdars,
  ``Thz imaging and sensing for security applications- explosives, weapons and
  drugs,'' \emph{Secmiconductor Science and Technology}, vol.~20, pp. 266--280,
  2005.

\bibitem{rogalski2003infrared}
A.~Rogalski, ``Infrared detectors: status and trends,'' \emph{Progress in
  quantum electronics}, vol.~27, no.~2, pp. 59--210, 2003.

\bibitem{lubecke98}
V.~Lubecke, K.~Mizuno, and G.~Rebeiz, ``Micromachining for terahertz
  applications,'' \emph{Microwave Theory and Techniques, IEEE Transactions on},
  vol.~46, no.~11, pp. 1821--1831, 1998.

\bibitem{Oliner59}
A.~Oliner and A.~Hessel, ``{Guided waves on sinusoidally-modulated reactance
  surfaces},'' \emph{{IRE Trans. Antennas Propag.}}, vol. AP-7, pp. 201--208,
  1959.

\bibitem{Grbic11_LWA_Sinusoidal}
A.~M. Patel and A.~Grbic, ``{A Printed Leaky-Wave Antenna Based on a
  Sinusoidally-Modulated Reactance Surface},'' \emph{{IEEE Trans. Antennas
  Propag.}}, vol.~59, no.~6, pp. 2087--2096, 2011.

\bibitem{Maci11}
S.~Maci, G.~Minatti, M.~Casaletti, and M.~Bosiljevac, ``Metasurfing: Addressing
  waves on impenetrable metasurfaces,'' \emph{IEEE Antennas and Wireless
  Propagation Letters}, vol.~10, pp. 1499--1502, 2011.

\bibitem{Geim2007}
K.~Geim and K.~S. Novoselov, ``The rise of graphene,'' \emph{Nature materials},
  vol.~6, pp. 183--91, 2007.

\bibitem{Oliner_2007}
A.~A. Oliner and D.~R. Jackson, ``Leaky-wave antennas,'' in \emph{Antenna
  Engineering Handbook}, 4th~ed., J.~L. Volakis, Ed.\hskip 1em plus 0.5em minus
  0.4em\relax New York: McGraw-Hill, 2007.

\bibitem{Gusynin09}
V.~P. Gusynin, S.~G. Sharapov, and J.~B. Carbotte, ``On the universal ac
  optical background in graphene,'' \emph{New J. Physics}, vol.~11, p. 095013,
  2009.

\bibitem{Hanson09}
G.~W. Hanson, ``Dyadic green's functions for an anisotropic non-local model of
  biased graphene,'' \emph{IEEE Transactions on Antennas and Propagation},
  vol.~56, no.~3, pp. 747--757, March 2008.

\bibitem{quantumC07}
T.~Fang, A.~Konar, H.~Xiang, and D.~Jena, ``Carrier statistics and quantum
  capacitance of graphene sheets and ribbons,'' \emph{Appl. Phys. Lett.},
  vol.~91, p. 092109, 2007.

\bibitem{Maci11_Spiral_LWA}
G.~Minatti, F.~Caminita, M.~Casaletti, and S.~Maci, ``{Spiral Leaky-Wave
  Antennas Based on Modulated Surface Impedance},'' \emph{{IEEE Trans. Antennas
  Propag.}}, vol.~59, no.~12, 2011.

\bibitem{Itoh89}
T.Itoh, \emph{Numerical Techniques for Microwave and Millimeter-Wave Passive
  Structures}.\hskip 1em plus 0.5em minus 0.4em\relax Wiley-Interscience, 1989.

\bibitem{Pryputniewicz2002}
R.~J. Pryputniewicz, \emph{MEMS SUMMiTV technology}.\hskip 1em plus 0.5em minus
  0.4em\relax Worcester Polytechnic Institute, Worcester, MA, 2002.

\bibitem{Sievenpiper10_Holography}
B.~H. Fong, J.~S. Colburn, J.~J. Ottusch, J.~L. Visher, and D.~F. Sievenpiper,
  ``{Scalar and Tensor Holographic Artificial Impedance Surfaces},''
  \emph{{IEEE Trans. Antennas Propag.}}, vol.~58, no.~10, pp. 3212--3221, 2010.

\bibitem{Naftaly07_THz_TDS}
M.~Naftaly and R.~E. Miles, ``{Terahertz Time-Domain Spectroscopy for Material
  Characterization},'' \emph{{Proceedings of the IEEE}}, vol.~95, no.~8, pp.
  1658--1665, 2007.

\bibitem{Dean10}
C.~R. Dean, A.~F. Young, I.~Meric, C.~Lee, L.~Wang, S.~Sorgenfrei, K.~Watanabe,
  T.~Taniguchi, P.~Kim, K.~L. Shepard, and J.~Hone, ``Boron nitride substrates
  for high-quality graphene electronics,'' \emph{Nature Nanotech.}, no.~5, pp.
  722--726, 2010.

\bibitem{Bolotin08}
K.~Bolotin, K.~Sikes, Z.~Jiang, M.~Klima, G.~Fudenberg, J.~Hone, P.~Kima, and
  H.~Stormer, ``Ultrahigh electron mobility in suspended graphene,''
  \emph{Solid State Communications}, vol. 146, pp. 351--355, March 2008.

\bibitem{Nikitin11}
A.~Y. Nikitin, F.~Guinea, F.~J. Garc\'ia-Vidal, and L.~Mart\'in-Moreno, ``Edge
  and waveguide terahertz surface plasmon modes in graphene microribbons,''
  \emph{Phys. Rev. B}, vol.~84, p. 161407, 2011.

\bibitem{GomezTornero12}
J.~L. G\'omez-Tornero, G.~Goussetis, and A.\'Alvarez-Melc\'on, ``Correction of
  dielectric losses in practical leaky-wave antenna designs,'' \emph{Journal of
  Electromagnetic Waves and Applications}, vol.~21, no.~8, pp. 1025--1036,
  2007.

\bibitem{Woo10}
I.~Woo, T.~K. Nguyen, H.~Han, H.~Lim, and I.~Park, ``Four-leaf-clover-shaped
  antenna for a thz photomixer,'' \emph{Opt. Express}, vol.~18, no.~18, pp.
  18\,532--18\,542, Aug 2010.

\bibitem{Yi11}
N.~Khiabani, H.~Yi, and S.~Yaochun, ``Thz photoconductive antennas in pulsed
  systems and cw systems,'' \emph{International Workshop on Antenna Technology
  (iWAT)}, 2011.

\bibitem{pads12}
S.~H. Lee, M.~Choi, T.-T. Kim, S.~Lee, M.~Liu, X.~Yin, H.~K. Choi, S.~S. Lee,
  C.-G. Choi, S.-Y. Choi, X.~Zhang, and B.~Min, ``Switching terahertz waves
  with gate-controlled active graphene metamaterials,'' \emph{Nature
  Materials}, vol.~11, pp. 936--941, 2012.

\bibitem{fringing_13}
E.~Forati and G.~W. Hanson, ``Soft-boundary graphene nanoribbon formed by a
  graphene sheet above a perturbed ground plane: conductivity pro?le and spp
  modal current distribution,'' \emph{J. Opt.}, vol.~15, p. 114006, 2013.

\bibitem{08_Power_Graphene}
S.~Ghosh, I.~Calizo, D.~Teweldebrhan, E.~P. Pokatilov, D.~L. Nika, A.~A.
  Balandin, W.~B. dn~F.~Miao, and C.~N. Lau, ``Extremely high thermal
  conductivity of graphene: Prospects for thermal management applications in
  nanoelectronic circuits,'' \emph{Appl. Phys. Lett.}, vol.~92, p. 151911,
  2008.

\end{thebibliography}
\end{document}